\documentclass[twocolumn,showpacs,amsmath,amssymb,aps,pra,floatfix,superscriptaddress,10pt]{revtex4-2}

\usepackage[latin1]{inputenc}
\usepackage[american]{babel}
\usepackage[T1]{fontenc}
\usepackage{amsmath}
\usepackage{upgreek}
\usepackage{mhchem}
\usepackage[dvipsnames]{xcolor}
\usepackage{graphicx}
\usepackage{multirow}
\usepackage{physics}
\usepackage{changes}
\usepackage[caption=false]{subfig}
\usepackage{mathtools}

\usepackage{tikz}
\usetikzlibrary{decorations.pathmorphing}
\usetikzlibrary{shapes}
\usetikzlibrary{shapes.geometric}
\usepackage{tikz-feynman}
\usetikzlibrary{backgrounds}
\newcommand{\brm}[1]{\boldsymbol{\mathrm{#1}}}

\begin{document}

\title{Full leading-order nuclear polarization in highly charged ions}

\date{\today}

\author{Igor~A.~Valuev}
\email[Email: ]{igor.valuev@mpi-hd.mpg.de} 
\affiliation{Max-Planck-Institut f\"{u}r Kernphysik, Saupfercheckweg 1, 69117 Heidelberg, Germany}

\author{Natalia~S.~Oreshkina}
\email[Email: ]{natalia.oreshkina@mpi-hd.mpg.de} 
\affiliation{Max-Planck-Institut f\"{u}r Kernphysik, Saupfercheckweg 1, 69117 Heidelberg, Germany}

\begin{abstract}
The nuclear-polarization corrections to the energy levels of highly charged ions are systematically investigated to leading order in the fine-structure constant.
To this end, the notion of effective photon propagators with nuclear-polarization insertions is employed, where the nuclear excitation spectrum is calculated by means of the Hartree-Fock-based random-phase approximation.
The effective Skyrme force is used to describe the interaction between nucleons, and the model dependence is analyzed.
To leading order, the formalism predicts two contributions given by the effective vacuum-polarization and self-energy diagrams.
The existing ambiguity around the vacuum-polarization term is resolved by demonstrating that it is effectively absorbed in the standard finite-nuclear-size correction.
The self-energy part is evaluated with the full electromagnetic electron-nucleus interaction taken into account, where the importance of the effects of the nuclear three-currents is emphasized.
\end{abstract}

\maketitle

\section{Introduction}
Ever since the first experimental detection of a nuclear-structure effect (i.e., the field shift) in the hyperfine spectra of the thallium isotopes $^{203}\text{Tl}$ and $^{205}\text{Tl}$ in~1931~\cite{1931_Schueler}, there have been extraordinary advances in high-precision atomic physics, both from the experimental and the theoretical sides.
On the one hand, this has led to a multitude of remarkable applications such as stringent tests of quantum electrodynamics~(QED)~\cite{2003_Draganic, 2005_Gumberidze, 2017_Ullmann,2023_Morgner}, determination of fundamental physical constants~\cite{2014_Sturm, 2006_Shabaev_spec_dif, 2016_Yerokhin_alpha, 2020_Cakir}, including the search for their variation~\cite{2005_Andreev, 2010_Berengut, 2015_Windberger, 2017_Oreshkina, 2019_Bekker}, as well as the quest for physics beyond the Standard Model~\cite{2020_Counts, 2022_Ono, 2022_Debierre}.
On the other hand, these applications are now becoming more and more limited by the ability to accurately take into account the increasingly more pronounced nuclear-structure effects.
This is especially true for highly charged ions, which hold a special place in the field due to their simplicity and the extreme electromagnetic environments they provide.
These advantages, however, come with the price of strong sensitivity to the size~\cite{2020_Valuev, 2023_Xie}, the shape~\cite{2008_Kozhedub, 2019_Michel, 2023_Sun} and even the internal dynamic structure~\cite{Plunien_1989, 1991_Plunien, 1995_Plunien} of the nucleus.
The latter is known as the nuclear-polarization~(NP) effect, and it poses one of the biggest challenges in theoretical description of the spectra of highly charged ions.

On the fundamental level, the NP effect stems from the difference between a static nuclear charge distribution, which is a common approximation in atomic calculations, and the true dynamic nature of the nucleus with the intricate motion of individual protons and neutrons within it.
Thus, the first calculations of the NP corrections to atomic energy levels of hydrogenlike ions were carried out by means of the ordinary second-order perturbation theory~\cite{1984_Hoffmann}.
This approach, however, is immediately challenged for relativistic systems by the presence of the negative continuum in the electronic Dirac spectrum.
As with any other difficulties plaguing the relativistic single-particle quantum mechanics, this issue could only be resolved within a field-theoretical description~\cite{Plunien_1989, 1991_Plunien, 1995_Plunien}, where the vacuum contribution arises naturally.
Indeed, as it turned out, a naive extension of the summation incorporates the negative part of the Dirac spectrum with a wrong expression in the energy denominator~\cite{1996_Nefiodov}, while omitting this contribution leads to a significant overestimation of the NP effect.

Despite the success and the power of the field-theoretical approach for NP, there are still aspects of it that have not received enough attention in the literature.
Firstly, to leading order in the fine-structure constant, this formalism leads to two terms corresponding to the effective self-energy and vacuum-polarization diagrams; however, only the former is usually considered, while the latter is omitted entirely, with only a few exceptions in the literature~\cite{1996_Labzowsky, Haga_mu_2002, Haga_mu_2007}.
Additionally, the question of possible double counting in the effective vacuum polarization, which was raised in Ref.~\cite{1996_Labzowsky}, needs further clarification.
Secondly, in the calculations of the effective self-energy contribution, only the longitudinal, or Coulomb, part of the electron-nucleus interaction is often taken into account, whereas it has been shown in Ref.~\cite{Haga_e_2002} that the transverse interaction is far from being negligible.
Furthermore, from the nuclear side of the computations, the common practice of employing simple phenomenological sum rules for estimating the NP contributions from giant resonances~\cite{Rinker_1978}, which are dominant in the case of spherical nuclei, does not allow to reliably assess the theoretical uncertainty of the NP corrections.
To the best of our knowledge, there is only one fully microscopic study of NP for electronic systems, which was carried out for hydrogenlike $^{208}\text{Pb}^{81+}$ in the framework of the random-phase approximation~\cite{Haga_e_2002}.
However, these calculations were performed only for one set of the Migdal-force parameters, and the nuclear model dependence was not explored.

The aim of this paper is to present a complete overview and analysis of the NP effect in highly charged ions to leading order in the fine-structure constant~$\alpha$.
That means considering both the self-energy and the vacuum-polarization effective diagrams, taking into account the full electromagnetic electron-nucleus interaction, incorporating a microscopic description of the nucleus as well as investigating the dominant theoretical uncertainty stemming from the nuclear model dependence.
For this purpose, we first outline the general field-theoretical formalism and the computational techniques from both atomic and nuclear parts of the calculations.
Then we demonstrate for both electronic and muonic systems that the effective vacuum polarization is already contained in the standard finite-nuclear-size effect and thus does not need to be included in the NP correction.
Finally, we present our results for the effective self-energy diagram for a series of hydrogenlike ions and a wide range of parametrizations of the effective Skyrme force, which is used in this work for nuclear description.
Relative contributions from different nuclear excitation modes, the importance of the transverse electron-nucleus interaction, and the model uncertainties are discussed.

Relativistic system of units ($\hbar = c = 1$) and Heaviside charge units ($\alpha = e^2/4\pi, e<0$) are used throughout the paper.
Four-vectors~($x$) are represented by regular typeface, while bold upright letters are used for three-vectors~($\brm{x}$), whose lengths are denoted by non-bold upright
letters ($|\brm{x}| \coloneqq \mathrm{x}$).

\section{Formalism} 
\subsection{Modified photon propagator}
The starting point of any perturbative calculation is the definition of a zeroth-order approximation.
In the case of bound atomic systems, we begin with the following Lagrangian density~$\mathcal{L}^{(0)}$ (omitting the issue of renormalization in this brief overview for simplicity)~\cite{1996_Greiner_field, 1996_Weinberg}:
\begin{align}
\label{eq:L0}
\mathcal{L}^{(0)} = \mathcal{L}^{\text{free}}_{\text{EM}} + \hat{\bar{\psi}} \left( i\gamma^{\mu} \partial_{\mu} - m_{\text{e}} - e \gamma^{\mu} \mathcal{A}^{\text{stat}}_{\mu} \right) \hat{\psi},
\end{align}
where $\hat{\psi}(x)$ is the electron-positron field operator, $m_{\text{e}}$ is the electron mass, and $\gamma^{\mu}$ are the Dirac matrices.
$\mathcal{A}^{\mu}_{\text{stat}}(\brm{x})$ denotes a classical time-averaged electromagnetic four-potential generated by the nucleus, while $\mathcal{L}^{\text{free}}_{\text{EM}}$ is the Lagrangian density for the free electromagnetic field operator $\hat{A}^{\mu}_{\text{free}}(x)$~\cite{1996_Greiner_field}:
\begin{align}
\mathcal{L}^{\text{free}}_{\text{EM}} = -\dfrac{1}{4} \hat{F}_{\mu \nu} \hat{F}^{\mu \nu} - \dfrac{1}{2} \zeta \left( \partial_{\sigma} \hat{A}^{\sigma}_{\text{free}} \right)^2,
\end{align}
where $\hat{F}^{\mu \nu}(x) = \partial^{\mu} \hat{A}^{\nu}_{\text{free}}(x) - \partial^{\nu} \hat{A}^{\mu}_{\text{free}}(x)$, the second gauge-fixing term corresponds to the Lorenz condition, and the value of the parameter~$\zeta$ determines the gauge choice.
In this subsection, we assume for simplicity~$\zeta=1$.
The introduction of the field $\mathcal{A}^{\mu}_{\text{stat}}(\brm{x})$ in Eq.~\eqref{eq:L0} corresponds to the external field approximation~\cite{1996_Weinberg} and allows to describe bound states, which are of central interest in atomic physics.
The usual interaction term is given by~\cite{1996_Greiner_field}
\begin{align}
\label{eq:L_int}
\mathcal{L}_{\text{int}} = -e \hat{\bar{\psi}} \gamma^{\mu} \hat{\psi} \hat{A}^{\text{free}}_{\mu},
\end{align}
and it is responsible for all kinds of QED corrections, which can be calculated in the so-called Furry picture~\cite{1951_Furry} while treating the nucleus simply as a static charge distribution.

In order to treat the NP effect on the same field-theoretical footing, the total nuclear four-current density operator~$\hat{J}^{\mu}_{\text{N}}$ is introduced as the following sum~\cite{Plunien_1989}:
\begin{align}
\label{eq:J_nucl}
\hat{J}^{\mu}_{\text{N}}(x) = J^{\mu}_{\text{N},\,\text{stat}}(\brm{x}) + \hat{J}^{\mu}_{\text{N},\,\text{fluc}}(x),
\end{align}
with the classical static part $J^{\mu}_{\text{N},\,\text{stat}}$ corresponding to the average over the nuclear ground state and the fluctuating part $\hat{J}^{\mu}_{\text{N},\,\text{fluc}}$ describing the intrinsic nuclear dynamics.
In the same way as $J^{\mu}_{\text{N},\,\text{stat}}$ is taken into account by introducing the corresponding classical field $\mathcal{A}^{\mu}_{\text{stat}}$ in Eq.~\eqref{eq:L0}, a second-quantized photon field $\hat{A}^{\mu}_{\text{fluc}}$ can be associated with the fluctuating current $\hat{J}^{\mu}_{\text{N},\,\text{fluc}}$.
In this view, a bound atomic electron interacts with the total electromagnetic field
\begin{align}
\begin{split}
\hat{A}^{\mu}_{\text{total}}(x) & = \mathcal{A}^{\mu}_{\text{stat}}(\brm{x}) + \hat{A}^{\mu}_{\text{fluc}}(x) + \hat{A}^{\mu}_{\text{free}}(x) \\
& = \mathcal{A}^{\mu}_{\text{stat}}(\brm{x}) + \hat{A}^{\mu}_{\text{rad}}(x),
\end{split}
\end{align}
where the total quantum radiation field $\hat{A}^{\mu}_{\text{rad}}$ is defined as the sum of the free photon field $\hat{A}^{\mu}_{\text{free}}$ and the fluctuating part $\hat{A}^{\mu}_{\text{fluc}}$ generated by $\hat{J}^{\mu}_{\text{N},\,\text{fluc}}$.
The latter is described by the following equation of motion:
\begin{align}
\partial^2 \hat{A}^{\mu}_{\text{fluc}}(x) = \hat{J}^{\mu}_{\text{N},\,\text{fluc}}(x).
\end{align}
As a consequence, in order to keep utilizing the standard machinery of Wick's theorem and Feynman diagrams, one is led to the modified photon propagator
\begin{align}
\label{eq:mod_prop}
i\mathcal{D}_{\mu\nu}(x,x') & = \langle 0|T[\hat{A}^{\text{rad}}_{\mu}(x) \hat{A}^{\text{rad}}_{\nu}(x')]|0 \rangle, 
\end{align}
instead of the usual free photon propagator given by
\begin{align}
iD_{\mu\nu}(x-x') = \langle 0|T[\hat{A}^{\text{free}}_{\mu}(x) \hat{A}^{\text{free}}_{\nu}(x')]|0 \rangle,
\end{align}
where~$|0\rangle$ denotes the ``vacuum'' state in the presence of the external field~$\mathcal{A}^{\mu}_{\text{stat}}(\brm{x})$, which corresponds to the nucleus being in its ground state.
Since the mixed terms arising from Eq.~\eqref{eq:mod_prop} vanish,
only the $\langle 0|T[\hat{A}^{\text{fluc}}_{\mu}(x) \hat{A}^{\text{fluc}}_{\nu}(x')]|0 \rangle$ term needs to be evaluated.

By using the fact that the free photon propagator is the Green's function of the free equation of motion:
\begin{align}
\partial^2 D_{\mu\nu}(x) = \eta_{\mu\nu} \delta^{(4)}(x),
\end{align}
and performing integration by parts with vanishing boundary terms, one can show that
\begin{widetext}
\begin{align}
\label{eq:mod_prop_der1}
\begin{split}
\langle 0|T[\hat{A}^{\text{fluc}}_{\mu}(x) \hat{A}^{\text{fluc}}_{\nu}(x')]|0 \rangle & = \int d^4 x_1 \, d^4 x_2 \, \eta_{\mu\xi} \delta^{(4)}(x-x_1) \langle 0|T[\hat{A}^{\xi}_{\text{fluc}}(x_1) \hat{A}^{\zeta}_{\text{fluc}}(x_2)]|0 \rangle \eta_{\zeta\nu} \delta^{(4)}(x_2-x') \\
& = \int d^4 x_1 \, d^4 x_2 \, \{\partial^2_{x_1} D_{\mu \xi}(x-x_1)\} \langle 0|T[\hat{A}^{\xi}_{\text{fluc}}(x_1) \hat{A}^{\zeta}_{\text{fluc}}(x_2)]|0 \rangle \{\partial^2_{x_2} D_{\zeta \nu}(x_2-x')\} \\
& = \int d^4 x_1 \, d^4 x_2 \, D_{\mu \xi}(x-x_1) \langle 0|\partial^2_{x_2} \partial^2_{x_1} T[\hat{A}^{\xi}_{\text{fluc}}(x_1) \hat{A}^{\zeta}_{\text{fluc}}(x_2)]|0 \rangle D_{\zeta \nu}(x_2-x').
\end{split}
\end{align}
\end{widetext}
It is important to note that the derivatives in the last line of Eq.~\eqref{eq:mod_prop_der1} act not only on the fields $\hat{A}^{\mu}_{\text{fluc}}$ but also on the $\theta$-functions from the time-ordered product: 
\begin{align}
\begin{split}
T[\hat{A}(x_1) \hat{B}(x_2)] & = \theta(x_1^0-x_2^0)\hat{A}(x_1) \hat{B}(x_2) \\
& + \theta(x_2^0-x_1^0)\hat{B}(x_2) \hat{A}(x_1), 
\end{split}
\end{align}
producing an additional term that we denote as $iS^{\xi \zeta}_{\text{N}}(x_1,x_2)$:
\begin{align}
\begin{split}
& \langle 0|\partial^2_{x_2} \partial^2_{x_1} T[\hat{A}^{\xi}_{\text{fluc}}(x_1) \hat{A}^{\zeta}_{\text{fluc}}(x_2)]|0 \rangle \\
& = \langle 0|T[\hat{J}^{\xi}_{\text{N},\,\text{fluc}}(x_1) \hat{J}^{\zeta}_{\text{N},\,\text{fluc}}(x_2)]|0 \rangle + iS^{\xi \zeta}_{\text{N}}(x_1,x_2),
\end{split}
\end{align}
while the resulting two-point current correlation function defines the so-called NP tensor
\begin{align}
\label{eq:np_tensor}
i\Pi^{\xi \zeta}_{\text{N}}(x_1,x_2) = \langle 0|T[\hat{J}^{\xi}_{\text{N},\,\text{fluc}}(x_1) \hat{J}^{\zeta}_{\text{N},\,\text{fluc}}(x_2)]|0 \rangle.
\end{align} 
Going back to Eq.~\eqref{eq:mod_prop}, the expression for the modified photon propagator can be written as
\begin{align}
\label{eq:mod_prop_final_1}
\mathcal{D}_{\mu\nu}(x,x') & = D_{\mu\nu}(x-x') + D^{\text{NP}}_{\mu\nu}(x,x'),
\end{align}
defining the NP correction $D^{\text{NP}}_{\mu\nu}(x,x')$ to the free photon propagator as follows:
\begin{align}
\label{eq:mod_prop_final_2}
\begin{split}
& D^{\text{NP}}_{\mu\nu}(x,x') = \int d^4 x_1 \, d^4 x_2 \, D_{\mu \xi}(x-x_1) \\
& \times \left[ \Pi^{\xi \zeta}_{\text{N}}(x_1,x_2) + S^{\xi \zeta}_{\text{N}}(x_1,x_2) \right] D_{\zeta \nu}(x_2-x').
\end{split}
\end{align}

In the original paper~\cite{Plunien_1989} by Plunien~\textit{et al.}, where the modified photon propagator~\eqref{eq:mod_prop} was first introduced, only the NP tensor appeared in the final expression. 
However, it was pointed out back in 1961 by Kenneth Johnson that a time-ordered product of two currents is in general not a covariant function~\cite{1961_Johnson}.
Thus, the role of the $iS^{\xi \zeta}_{\text{N}}$ term is to maintain the Lorentz covariance of the vacuum expectation value~$\langle 0|T[\hat{A}^{\text{fluc}}_{\mu}(x) \hat{A}^{\text{fluc}}_{\nu}(x')]|0 \rangle$ and the modified photon propagator as a whole.
In addition, Lowell Brown demonstrated a connection between the requirement of gauge invariance and the restored Lorenz covariance of a properly defined two-point current correlation operator~\cite{1966_Brown}.
This requirement implies that
\begin{align}
\partial_{x_1,\xi} \left( i\Pi^{\xi \zeta}_{\text{N}}(x_1,x_2) + iS^{\xi \zeta}_{\text{N}}(x_1,x_2) \right) = 0,
\end{align}
which leads to
\begin{align}
\label{eq:schwinger_seagull}
\begin{split}
\langle 0|\delta(x^0_1-x^0_2)[\hat{J}^{0}_{\text{N},\,\text{fluc}}(x_1), \hat{J}^{\zeta}_{\text{N},\,\text{fluc}}(x_2)]| 0\rangle& \\ 
+\hspace{2pt} \langle 0|T[\partial_{x_1,\xi}\hat{J}^{\xi}_{\text{N},\,\text{fluc}}(x_1) \hat{J}^{\zeta}_{\text{N},\,\text{fluc}}(x_2)]|0\rangle& \\ 
+\hspace{2pt} i\partial_{x_1,\xi}S^{\xi \zeta}_{\text{N}}(x_1,x_2)& = 0,
\end{split}
\end{align}
where the second term is equal to zero due to the continuity equation of nuclear charge conservation.
While the equal-time commutator in Eq.~\eqref{eq:schwinger_seagull} vanishes for \mbox{$\zeta=0$}, it was shown by Julian Schwinger from fundamental principles of quantum field theory that charge and current~($\zeta=1,2,3$) densities cannot commute at a common time~\cite{1959_Schwinger}.
It follows from Eq.~\eqref{eq:schwinger_seagull} that these non-vanishing commutators, known as the Schwinger terms, must be cancelled by the divergence of~$iS^{\xi\zeta}_{\text{N}}$, if gauge invariance is to be satisfied.
The contribution~$iS^{\xi\zeta}_{\text{N}}$ is often called the ``seagull'' or ``catastrophic'' term, and this kind of cancellation is in fact a very general result in current-algebra theories~\cite{1967_Brown}.

It is clear that the expression for~$iS^{\xi\zeta}_{\text{N}}$ depends on a specific definition of $\hat{J}^{\mu}_{\text{N}}$.
For example, it can be shown that in the case of the non-relativistic nuclear charge-current density operators the seagull term takes on the following form~\cite{Haga_e_2002}:
\begin{align}
\label{eq:seagull_nonrel}
S^{\xi \zeta}_{\text{N}}(x_1,x_2) & = \dfrac{|e|\langle 0|\hat{\rho}_{\text{N}}(\brm{x}_1)|0\rangle}{M_{\text{p}}} \delta^{\xi \zeta} \delta^{(4)}(x_1-x_2), 
\end{align}
where $\hat{\rho}_{\text{N}}(\brm{x}_1)\coloneqq \hat{J}^{0}_{\text{N}}(0,\brm{x}_1)$, $M_{\text{p}}$ is the proton mass, and $\delta^{\xi \zeta}$ is the Kronecker delta extended to four dimensions with $\delta^{00}=0$.

\subsection{Nuclear-polarization insertion}
As one can see from Eq.~\eqref{eq:mod_prop_final_1}, every photon line in ordinary QED diagrams receives the NP correction in the form of Eq.~\eqref{eq:mod_prop_final_2}.
The part $[\Pi^{\xi \zeta}_{\text{N}} + S^{\xi \zeta}_{\text{N}}]$ in between the two free photon propagators is called the NP insertion, and it contains all the information about the intrinsic nuclear dynamics.
In order to implement this correction in practical calculations, we employ the two-time Green's function method from Ref.~\cite{2002_Shabaev}.
For this purpose, we need a Fourier-transformed version of~$D^{\text{NP}}_{\mu\nu}$ with respect to the time variables.

First, by writing the time evolution of the Heisenberg operators~$\hat{J}^{\mu}_{\text{N},\,\text{fluc}}(t,\brm{x})$, inserting a complete set of nuclear excitations $|\lambda\rangle$ in Eq.~\eqref{eq:np_tensor} (the ground state does not contribute due to the definition in Eq.~\eqref{eq:J_nucl}) and using the integral representation of the $\theta$\nobreakdash-function, it is easy to show that the NP tensor is homogeneous in time~\cite{1991_Plunien}:
\begin{align}
\label{eq:ft_np_tensor}
\begin{split}
\Pi^{\xi \zeta}_{\text{N}} (t_1-t_2, \boldsymbol{\mathrm{x}}_1, \boldsymbol{\mathrm{x}}_2) = \int \dfrac{d \omega}{2\pi} \, &e^{-i\omega \left( t_1 - t_2 \right)} \, \\
&\times \widetilde{\Pi}^{\xi \zeta}_{\text{N}} (\omega, \boldsymbol{\mathrm{x}}_1, \boldsymbol{\mathrm{x}}_2), 
\end{split} 
\end{align}
with
\begin{align}
\label{eq:np_tensor_omega}
\begin{split}
\widetilde{\Pi}^{\xi \zeta}_{\text{N}} (\omega, \brm{x}_1, \brm{x}_2) = \sum_{\lambda} &\left( \dfrac{\langle 0|\hat{J}^{\xi}_{\text{N}}\left(\brm{x}_1\right)|\lambda \rangle \langle \lambda|\hat{J}^{\zeta}_{\text{N}}\left(\brm{x}_2\right)|0 \rangle}{\omega - \omega_{\lambda} + i0} \right. \\
&\left. - \dfrac{\langle \lambda|\hat{J}^{\xi}_{\text{N}}\left(\brm{x}_1\right)|0 \rangle \langle 0|\hat{J}^{\zeta}_{\text{N}}\left(\brm{x}_2\right)|\lambda \rangle}{\omega + \omega_{\lambda} - i0} \right).
\end{split}
\end{align}
where $\hat{J}^{\mu}_{\text{N}}(\brm{x})\coloneqq \hat{J}^{\mu}_{\text{N}}(0,\brm{x})$, and $\omega_{\lambda} = E_{\lambda} - E_0$ are the nuclear excitation energies.
Similarly, the Fourier-transformed version of the seagull term~\eqref{eq:seagull_nonrel} reads
\begin{align}
\label{eq:seagull_ft}
\widetilde{S}^{\xi \zeta}_{\text{N}}(\omega,\brm{x}_1,\brm{x}_2) = \dfrac{|e|\langle 0|\hat{\rho}_{\text{N}}(\brm{x}_1)|0\rangle}{M_{\text{p}}} \delta^{\xi \zeta} \delta^{(3)}(\brm{x}_1-\brm{x}_2).
\end{align}
Then, by using Eq.~\eqref{eq:ft_np_tensor} together with
\begin{align}
D_{\mu \xi} \left( x-x_1 \right) = \int \dfrac{d\omega}{2\pi} \, e^{-i\omega \left( t - t_1 \right)} \widetilde{D}_{\mu \xi} \left( \omega, \brm{x} - \brm{x}_1 \right),
\end{align}
one readily obtains the desired equivalent of Eq.~\eqref{eq:mod_prop_final_2}:
\begin{align}
&\widetilde{D}^{\text{NP}}_{\mu\nu}(\omega,\brm{x},\brm{x}') = \int d^3 x_1 d^3 x_2 \, \widetilde{D}_{\mu\xi}(\omega,\brm{x}-\brm{x}_1) \\
&\times \left[ \widetilde{\Pi}^{\xi\zeta}_{\text{N}}(\omega,\brm{x}_1,\brm{x}_2) + \widetilde{S}^{\xi \zeta}_{\text{N}}(\omega,\brm{x}_1,\brm{x}_2) \right] \widetilde{D}_{\zeta\nu}(\omega,\brm{x}_2-\brm{x}'). \notag
\end{align}

\begin{figure}[!htbp]
\centering
\includegraphics[scale=1]{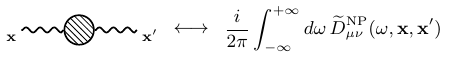}
\caption{Feynman rule for evaluating the NP correction to the photon propagator.}%
\label{fig:np_rule}
\end{figure}

Thus, we can supplement the set of the Feynman rules in Ref.~\cite{2002_Shabaev} with an additional one shown in Fig.~\ref{fig:np_rule}.
The NP insertion is represented by a shaded circle, and, by analogy with the rule for an internal photon line, the energy variable~$\omega$ has to be integrated over.

\begin{figure*}[!htbp]%
\centering
\subfloat[\centering]{{ \includegraphics[scale=1]{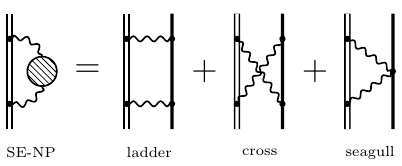} }}%
\qquad \qquad \qquad \qquad \qquad
\subfloat[\centering]{{ \includegraphics[scale=1]{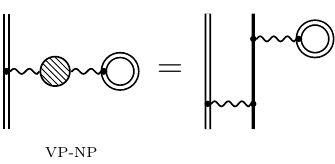} }}%
\vspace{-0.1cm}
\caption{Effective self-energy~(a) and vacuum-polarization~(b) diagrams as a result of applying the NP insertion (shaded circle) to the photon propagator (wavy line). The former can be interpreted as a two-photon exchange between the bound atomic electron (double line) and the nucleus (single solid line), which includes the ladder, the cross and the seagull diagrams; while the latter can be understood as the interaction of the induced vacuum polarization with the bound electron via additional virtual nuclear excitations.}%
\label{fig:diagrams}%
\end{figure*}

To leading order in the fine-structure constant, the NP insertion can be applied to both the self-energy~(SE) and the vacuum-polarization~(VP) corrections resulting in the two effective diagrams presented in Fig.~\ref{fig:diagrams}.
Analogously to the ordinary SE and VP effects, the following general expressions for the energy shifts of an electronic level~$|a\rangle$ due to the SE-NP and VP-NP diagrams are obtained as
\begin{align}
&\begin{aligned}
\label{eq:se_np}
\Delta E^{\text{SE-NP}}_{a} & = \dfrac{ie^2}{2\pi} \sum_{n} \int d^3 x d^3 x' d\omega \, \widetilde{D}^{\text{NP}}_{\mu\nu}(\omega,\brm{x},\brm{x}') \\
&\times \dfrac{\bar{\psi}_{a}(\brm{x})\gamma^{\mu}\psi_{n}(\brm{x}) \bar{\psi}_{n}(\brm{x}')\gamma^{\nu}\psi_{a}(\brm{x}')}{\varepsilon_a-\omega-\varepsilon_{n}(1-i0)},
\end{aligned} \\[5pt]
&\begin{aligned}
\label{eq:vp_np}
\Delta E^{\text{VP-NP}}_{a} & = -\dfrac{ie^2}{2\pi} \int d^3 x d^3 x' d\eta \, \widetilde{D}^{\text{NP}}_{\mu\nu}(0,\brm{x},\brm{x}') \\
&\times \bar{\psi}_{a}(\brm{x})\gamma^{\mu}\psi_{a}(\brm{x}) \mathrm{Tr}\left[ S(\eta,\brm{x}',\brm{x}') \gamma^{\nu} \right],
\end{aligned}
\end{align}
where the dressed electron propagator
\begin{align}
\label{eq:e_prop_omega}
S(\eta,\brm{x}_1,\brm{x}_2) = \sum_n \dfrac{\psi_n(\brm{x}_1)\bar{\psi}_n(\brm{x}_2)}{\eta-\varepsilon_n(1-i0)}
\end{align}
is constructed from a complete set of eigenstates and eigenvalues of the stationary single-particle Dirac equation:
\begin{align}
\label{eq:Dirac}
\left[ -i\boldsymbol{\alpha} \cdot \nabla + \beta m_{\text{e}} + V(\brm{x}) \right] \psi_n(\brm{x}) = \varepsilon_n \psi_n(\brm{x}),
\end{align}
with $V(\brm{x}) = e\mathcal{A}_0(\brm{x})$ denoting the potential energy corresponding to the electrostatic potential of an atomic nucleus, $\beta=\gamma^0$, and $\alpha^{i}=\gamma^{0}\gamma^{i}$ $(i=1,2,3)$.

The SE-NP contribution has received much more attention in the literature because of its correspondence to the usual NP correction calculated earlier in second-order perturbation theory.
In the language of field theory, Eq.~\eqref{eq:se_np} can also be interpreted as a two-photon exchange between the bound electron and the nucleus, with the ladder and the cross diagrams in Fig.~\ref{fig:diagrams}(a) corresponding to the first and the second terms in the expression~\eqref{eq:np_tensor_omega} for the NP tensor, respectively.
The seagull term~\eqref{eq:seagull_ft} can be represented as a coupling of the electromagnetic currents to the nucleus at the same point, and its physical significance is to ensure gauge invariance of the calculated SE-NP corrections.
In this work, we employ the formulas for the ladder, cross, and seagull terms in the momentum representation that can be found in Ref.~\cite{Haga_e_2002}.
As for the VP-NP contribution, we present some more details on it in the following subsection.
\subsection{Effective vacuum polarization} \label{vpnp}
The trace part of Eq.~\eqref{eq:vp_np} can be rewritten as~\cite{1985_Greiner_QED_strong}
\begin{align}
-ie \mathrm{Tr}\left[ \int \dfrac{d\eta}{2\pi} S(\eta,\brm{x}',\brm{x}') \gamma^{\nu} \right] & = -ie \mathrm{Tr}\left[ S(x,x') \gamma^{\nu} \right]_{x \underset{\text{sym}}{\to} x'}  \notag \\
& = \langle 0| \hat{j}^{\nu}(\brm{x}')|0 \rangle,
\end{align}
where the current operator of the Dirac field is defined as $\hat{j}^{\nu}(\brm{x}')=\frac{e}{2} [\hat{\bar{\psi}}(\brm{x}'), \gamma^{\nu}\hat{\psi}(\brm{x}')]$, and the limit~$x=x'$ is approached symmetrically from~$t=t'-0$ and~$t=t'+0$. 
To first order in~$Z\alpha$, this induced current is related to the static external current of the nucleus in momentum space as follows~\cite{1985_Greiner_QED_strong}:
\begin{align}
\label{eq:induced_current}
\langle \hat{j}^{\nu}(\brm{x}') \rangle^{(1)} = \int \dfrac{d^3 k}{(2\pi)^3} \Pi^{(1)}_{\text{ren}}(-\brm{k}^2) \widetilde{J}^{\nu}_{\text{N},\,\text{stat}}(\brm{k}) e^{i\brm{k}\cdot \brm{x}'},
\end{align}
with the renormalized Uehling polarization function
\begin{align}
\begin{split}
\Pi^{(1)}_{\text{ren}}(-\brm{k}^2) &= -\dfrac{2\alpha}{3 m_{\text{e}}\pi} \int_1^{\infty} \dfrac{d\xi}{\xi} \,  \sqrt{1 - \dfrac{1}{\xi^2}} \\
&\times \left( 1 + \dfrac{1}{2\xi^2} \right) \dfrac{\brm{k}^2}{\brm{k}^2 + (2m_{\text{e}}\xi)^2}.
\end{split}
\end{align}

Since we consider electrostatic nuclear potentials, the only non-vanishing component in Eq.~\eqref{eq:induced_current} is the nuclear ground-state charge density $\tilde{J}^{0}_{\text{N},\,\text{stat}}(\brm{k}) \coloneqq \tilde{\rho}_{\text{N}}(\brm{k})$.
As a result, only the component~$D^{\text{NP}}_{00}(0,\brm{x},\brm{x}')$ contributes to the VP\nobreakdash-NP energy shift:
\begin{align}
\label{eq:vp_np_2}
\begin{split}
\Delta E^{\text{VP-NP}(1)}_{a} = e \int d^3 x d^3 x' \, &\psi^{\dagger}_{a}(\brm{x}) \psi_{a}(\brm{x}) \\
\times &\widetilde{D}^{\text{NP}}_{00}(0,\brm{x},\brm{x}') \rho^{(1)}_{\text{VP}}(\brm{x}'),
\end{split}
\end{align}
where $\rho^{(1)}_{\text{VP}}(\brm{x}') \coloneqq \langle \hat{j}^{0}(\brm{x}') \rangle^{(1)}$ is the induced vacuum-polarization density.
In this case, the NP correction to the photon propagator is obtained most conveniently in the Coulomb gauge:
\begin{align}
\label{eq:D_00_1}
\begin{split}
\widetilde{D}^{\text{NP}}_{00}(0,\brm{x},\brm{x}') = \int &d^3 x_1 d^3 x_2 \, \dfrac{1}{4\pi|\brm{x}-\brm{x}_1|} \\
&\times \widetilde{\Pi}^{00}_{\text{N}}(0,\brm{x}_1,\brm{x}_2) \dfrac{1}{4\pi|\brm{x}_2-\brm{x}'|}.
\end{split}
\end{align}
Moreover, in the case of spherically symmetric nuclear charge distributions~$\tilde{\rho}_{\text{N}}$, the induced vacuum-polarization density $\rho^{(1)}_{\text{VP}}$ becomes spherical, too.
This fact leads to a further simplification of Eqs.~\eqref{eq:vp_np_2} and~\eqref{eq:D_00_1} when one takes into account the following expression for the nuclear matrix elements entering the NP tensor~\cite{skyrme_rpa}:
\begin{align}
\label{eq:td}
\langle \lambda(JM)|\hat{\rho}_{\text{N}}(\brm{x})|0 \rangle = \rho^{\text{tr}}_{\lambda}(\mathrm{x}) Y^{*}_{JM}(\Omega_{\brm{x}}),
\end{align}
where $\rho^{\text{tr}}_{\lambda}(\mathrm{x})$ is the radial charge transition density for a nuclear excited state~$|\lambda\rangle$ with the angular quantum numbers~$J$ and~$M$.
Thus, it is easy to see that only the monopole part of the Coulomb interaction in Eq.~\eqref{eq:D_00_1} survives, which also means that only the monopole nuclear excitations (the so-called breathing modes) contribute to the VP-NP correction.
In this case, the NP tensor takes on an especially simple form:
\begin{align}
\widetilde{\Pi}^{00}_{\text{N}}(0,\mathrm{x}_1,\mathrm{x}_2) = -\sum_{\lambda(J=0)} \dfrac{1}{2\pi\omega_{\lambda}} \rho^{\text{tr}}_{\lambda}(\mathrm{x}_1) \rho^{\text{tr}}_{\lambda}(\mathrm{x}_2),
\end{align}
such that
\begin{align}
\widetilde{D}^{\text{NP}}_{00}(0,\mathrm{x},\mathrm{x}') = -\sum_{\lambda(J=0)} \dfrac{1}{2\pi \omega_{\lambda}} I^{\text{tr}}_{\lambda}(\mathrm{x}) I^{\text{tr}}_{\lambda}(\mathrm{x}'),
\end{align}
where
\begin{align}
I^{\text{tr}}_{\lambda}(\mathrm{x}) = \int_{0}^{\infty} \mathrm{y}^2d\mathrm{y} \, \dfrac{\rho^{\text{tr}}_{\lambda}(\mathrm{y})}{\mathrm{max}(\mathrm{x},\mathrm{y})}.
\end{align}
Finally, we also recall that in central potentials~$V(\mathrm{x})$ the electron wave function factorizes into the radial and angular parts:
\begin{align}
\psi_{n\kappa m}(\brm{x}) = \dfrac{1}{\mathrm{x}} 
   \begin{pmatrix}
   G_{n\kappa}(\mathrm{x})\mathit{\Omega}_{\kappa m}(\Omega_{\brm{x}}) \\[3pt]
   iF_{n\kappa}(\mathrm{x})\mathit{\Omega}_{-\kappa m}(\Omega_{\brm{x}})
   \end{pmatrix},
\end{align}
where $n$ is the principal quantum number, $\kappa$ is the relativistic angular momentum number, $m$ is the total magnetic number, and $\mathit{\Omega}_{\pm\kappa m}(\Omega_{\brm{x}})$ are the spherical spinors.
After collecting everything together, we obtain the following expression for the VP-NP energy shift:
\begin{align}
\Delta E^{\text{VP-NP}(1)}_{a} = -e \sum_{\lambda(J=0)} \dfrac{1}{2\pi\omega_{\lambda}} M_{\lambda,a} I^{\text{VP}}_{\lambda},
\end{align}
with
\begin{align}
M_{\lambda,a} &= \int_{0}^{\infty} d\mathrm{x} \left[ G^2_{n_a \kappa_a}(\mathrm{x}) + F^2_{n_a \kappa_a}(\mathrm{x}) \right] I^{\text{tr}}_{\lambda}(\mathrm{x}), \\
I^{\text{VP}}_{\lambda} &= 4\pi \int_{0}^{\infty} \mathrm{x}^2d\mathrm{x} \, \rho^{(1)}_{\text{VP}}(\mathrm{x}) I^{\text{tr}}_{\lambda}(\mathrm{x}).
\end{align}
We note that the VP-NP correction can also be expressed in terms of the reduced transition probabilities~$B(E0)$, which was done in Ref.~\cite{1996_Labzowsky}.
\subsection{Computational techniques}
The calculations of the NP corrections require input from both atomic and nuclear physics.
For the atomic part, we solve the stationary Dirac equation~\eqref{eq:Dirac} numerically by confining the system to a spherical cavity of a large radius and expanding the radial part of the electron wavefunction in terms of \textit{B}\nobreakdash-splines within the dual-kinetic-balance approach~\cite{2004_Shabaev_DKB}. 
The resulting generalized matrix eigenvalue equations are then readily solved by means of the LAPACK library.
This approach is especially useful in the case of the SE-NP correction since it allows to reduce the infinite sum over the bound states and the integrals over the positive and negative continua to finite sums with no remainders to evaluate.
At the same time, the convergence of the results is readily controlled by varying the size of the cavity and the number of \textit{B}\nobreakdash-splines used.
As for the central nuclear potential~$V(\mathrm{x})$ corresponding to the zeroth approximation, it is sufficient to construct it from the simple two-parameter Fermi charge distribution $\rho_{\text{N}}(\mathrm{x}) = \rho_0 \{ 1 + \mathrm{exp} [(\mathrm{x}-c)/a] \}^{-1}$, where we use the standard value of the diffuseness parameter $a = 2.3/[4 \, \mathrm{ln}(3)] \, \mathrm{fm}$ and adjust the half-density radius $c$ such that the tabulated values~\cite{2013_Angeli} of the root-mean-square~(RMS) radii are reproduced.

For the nuclear part of the computations, we employ the~\verb|skyrme_rpa| program~\cite{skyrme_rpa} in order to obtain the entire nuclear excitation spectrum.
In the first step, the self-consistent Hartree-Fock mean field is built assuming the effective Skyrme-type interaction between the nucleons. 
Then, based on this mean field, the excited nuclear states are calculated in the framework of the random-phase approximation~(RPA).
Under the spherical symmetry assumption, the RPA equations are solved separately for given angular momentum and parity~$J^{\pi}$.
In our calculations we take into account the $0^{+}$, $1^{-}$, $2^{+}$, $3^{-}$, $4^{+}$, $5^{-}$, and $1^{+}$ excitation modes.
The completeness of the obtained spectra is controlled by the degree of fulfilment of the energy-weighted sum rule.
In addition, we extended the code to include the transition densities for the nuclear three-currents, which are needed for the transverse part of the SE-NP corrections.
Non-relativistic charge-current operators are used in the calculations of the nuclear matrix elements, and the corresponding expressions can be found in Ref.~\cite{Tanaka_1994}.
\section{Results and discussion}
\subsection{Effective vacuum polarization}
\begin{table}[!tbp]
\begin{center}
\caption{Comparison between the VP-NP corrections and the increase of the FNS effects stemming from the enlargement of the nuclear RMS charge radius (first row) due to introduction of the Uehling potential into the mean field of the protons.
The values are given for the $^{208}\text{Pb}$ nucleus with a single bound electron or muon in the $1s_{1/2}$, $2s_{1/2}$, and $2p_{1/2}$ states.
Three Skyrme parametrizations SKX, KDE0, and SkP are chosen as examples.}
{\renewcommand{\arraystretch}{1.2}
\renewcommand{\tabcolsep}{0.24cm}
\begin{tabular}{llccc}
\hline \hline
& & SKX & KDE0 & SkP \\
\hline
\multicolumn{2}{c}{$\Delta R_{\text{RMS}}$(fm)} & 0.00112 & 0.00128 & 0.00142 \\
\hline
\multirow{6}{*}{$e^{-}$(meV)} & $\Delta \text{FNS}_{1s_{1/2}}$ & 21.6 & 24.8 & 27.6 \\
& $\text{VP-NP}_{1s_{1/2}}$ & 21.1 & 23.6 & 26.9 \\
& $\Delta \text{FNS}_{2s_{1/2}}$ & 3.75 & 4.30 & 4.78 \\
& $\text{VP-NP}_{2s_{1/2}}$ & 3.65 & 4.10 & 4.66 \\
& $\Delta \text{FNS}_{2p_{1/2}}$ & 0.32 & 0.37 & 0.41 \\
& $\text{VP-NP}_{2p_{1/2}}$ & 0.32 & 0.35 & 0.40 \\
\hline
\multirow{6}{*}{$\mu^{-}$(keV)} & $\Delta \text{FNS}_{1s_{1/2}}$ & 1124 & 1289 & 1432 \\
& $\text{VP-NP}_{1s_{1/2}}$ & 1110 & 1247 & 1407 \\
& $\Delta \text{FNS}_{2s_{1/2}}$ & 218 &  250 & 277 \\
& $\text{VP-NP}_{2s_{1/2}}$ & 217 & 244 & 274 \\
& $\Delta \text{FNS}_{2p_{1/2}}$ & 156 & 179 & 199 \\
& $\text{VP-NP}_{2p_{1/2}}$ & 151 & 169 & 193 \\
\hline \hline
\end{tabular}} 
\label{tab:vpnp_fns}
\end{center}
\end{table}
In this section we present the numerical results of our NP calculations.
First, we address the VP-NP diagram and its physical meaning, as the question was raised in Ref.~\cite{1996_Labzowsky} on whether this contribution might already be contained in the ordinary finite-nuclear-size~(FNS) effect.
Based on a scaling argument, the authors of Ref.~\cite{1996_Labzowsky} suggested that there is a part of the VP-NP correction that is qualitatively distinct from the standard FNS contribution; however, they did not provide a clear procedure of extracting it. 
At the same time, the diagrammatic representation in Fig.~\ref{fig:diagrams}(b) suggests that the VP-NP effect corresponds to a correction to the Coulomb potential of the nucleus due to the interaction between the induced vacuum polarization and nuclear degrees of freedom.
Therefore, in this work, we approach this issue by additionally considering the effect of the induced vacuum polarization on the nuclear ground state.

To this end, we modify the \verb|skyrme_rpa| program by introducing the Uehling potential~\cite{1976_Fullerton}
\begin{align}
\begin{split}
V_{\text{Ue}}(\mathrm{r}) &= -\dfrac{2\alpha^2}{3 m_{\text{e}} \mathrm{r}} \int_{0}^{\infty} d\mathrm{r}' \, \mathrm{r}' \rho_{\text{p}}({\mathrm{r}'}) \int_{1}^{\infty} \dfrac{d\xi}{\xi^2} \, \sqrt{1 - \dfrac{1}{\xi^2}} \\
&\times \left( 1 + \dfrac{1}{2\xi^2} \right) \left( e^{-2m_{\text{e}} |\mathrm{r} - \mathrm{r}'| \xi} - e^{-2m_{\text{e}} (\mathrm{r} + \mathrm{r}') \xi} \right)
\end{split}
\end{align}
into the mean field generated by the protons with the density distribution~$\rho_{\text{p}}({\mathrm{r}'})$ and estimate the resulting change in the RMS charge radius.
The obtained enlargement of the nucleus can then be translated into the corresponding increase of the FNS effect (taken with respect to the tabulated RMS radius), which we denote as~$\Delta \text{FNS}$.
The VP\nobreakdash-NP corrections calculated from the monopole excitation spectrum as described in Subsection~\ref{vpnp} also decrease the atomic binding energies and can be directly compared to the $\Delta \text{FNS}$~values.
We present such a comparison in Table~\ref{tab:vpnp_fns} for the example of the $^{208}\text{Pb}$ nucleus and three parametrizations of the Skyrme force, namely SKX, KDE0, and SkP~\cite{SKX, KDE0, SkP}.
In order to examine the scaling behaviour, both the electron and the muon are considered as a bound atomic particle, and the results are given for the $1s_{1/2}$, $2s_{1/2}$, and $2p_{1/2}$ states.
Considering both leptons is of importance, given that the nuclear charge radii are determined experimentally from muonic spectroscopy~\cite{1995_Fricke} and subsequently used as input parameters in the description of electronic systems.
Therefore, it is essential to treat these two cases consistently.

The extremely close agreement between the $\Delta \text{FNS}$ and VP\nobreakdash-NP values, independent of the state and the bound particle considered, is clear evidence that the VP\nobreakdash-NP effect simply corresponds to the change in the nuclear charge radius due to the induced vacuum polarization.
Put differently, if one calculates the VP\nobreakdash-NP correction and expresses the corresponding energy shift in terms of a shift of the RMS radius, the resulting radius shift will be the same for different states and also for electronic and muonic atoms.
We note that a similar level of agreement was also found for other nuclei considered in this work.
Since the experimentally obtained nuclear charge radii are inseparable from such QED corrections and thus already contain them, we conclude that the VP\nobreakdash-NP contribution does not need to be taken into account as part of the NP effect.
\subsection{Effective self-energy in Coulomb approximation}\label{se_coulomb}
Before turning to the full SE-NP correction, let~us first consider the approximation commonly used in the literature.
Based on the argument that the velocities associated with nuclear dynamics are mainly non-relativistic, the contributions from the nuclear three-currents~$\hat{\brm{J}}_{\text{N}}$ are often neglected, meaning that only the~$\Pi^{00}_{\text{N}}$ component of the NP~tensor (the longitudinal, or Coulomb, part) is taken into account.
We refer to this framework as the ``Coulomb approximation''.
Moreover, the calculations can be simplified even further by expressing the SE-NP energy shift in terms of the experimentally measurable nuclear transition probabilities $B(EL; L \rightarrow 0)_{\lambda}$ and excitation energies~$\omega_{\lambda}$, resulting in the most widely used formula for the SE\nobreakdash-NP correction~\cite{1996_Nefiodov}:
\begin{align}
\label{eq:NP_Nef}
\Delta E^{\text{SE-NP}}_{a} \overset{\text{C}}{=} -\alpha \sum_{n,\lambda,L,M} \dfrac{B(EL)_{\lambda} |\langle n|F_L Y_{LM}|a \rangle|^2}{\varepsilon_n - \varepsilon_a + \mathrm{sgn}(\varepsilon_n)\omega_{\lambda}},
\end{align}
where the radial functions $F_L$ are often assumed to be
\begin{align}
&\begin{aligned}
\label{eq:F_0}
F_0(\mathrm{x}) & = \dfrac{5\sqrt{\pi}}{2R_0^3} \left[ 1 - \left( \dfrac{\mathrm{x}}{R_0} \right)^2 \right] \theta(R_0 - \mathrm{x}), 
\end{aligned}\\
&\begin{aligned}
\label{eq:F_L}
F_L(\mathrm{x}) = \dfrac{4\pi}{(2L+1)R_0^L} \biggl[ &\dfrac{\mathrm{x}^L}{R_0^{L+1}} \theta(R_0-\mathrm{x}) \\
+ &\dfrac{R_0^L}{\mathrm{x}^{L+1}} \theta(\mathrm{x}-R_0) \biggr], \quad L \geq 1,
\end{aligned}
\end{align}
with $R_0$ being the radius of the nucleus as a homogeneously charged sphere.
However, the experimental $B(EL)_{\lambda}$ and $\omega_{\lambda}$ values are available only for low-lying nuclear states, while the crucial contributions from the giant resonances have to be estimated by resorting to phenomenological energy-weighted sum rules~(EWSR).
In this approach, the giant resonances are assumed to be concentrated in a single state for each multipolarity and isospin.
Since the \verb|skyrme_rpa| program also allows to compute the~$B(EL)_{\lambda}$ values, it is of interest to compare the Coulomb SE-NP corrections calculated from such a microscopic nuclear theory with the results obtained from the experimental and EWSR nuclear parameters.
Such a comparison for the ground state of the $^{208}\text{Pb}^{81+}$ ion is presented in Table~\ref{tab:bel}.
Eleven different Skyrme parametrizations~\cite{KDE0, SKX, 1998_Chabanat, BSk14, SAMi, NRAPR, SkP, SkM, SGII, SKI3, LNS} (nine of which have been shown to cover a wide range in the parameter space~\cite{2022_Valuev}) have been used in the microscopic calculations, while the last row has been calculated using the same $B(EL)_{\lambda}$ and $\omega_{\lambda}$ values as in Ref.~\cite{1996_Nefiodov}.
The results turn out to be quite stable with respect to a nuclear model and are in remarkably good agreement with the extremely simple EWSR estimations.
To appreciate this fact, we note that, unlike concentrating the giant resonances in just a few ``states'' in the EWSR approach, the microscopic RPA calculations produce numerous excitations, e.g., around 1500 for the $3^{-}$ mode alone.
It is also immediately apparent that most of the nuclear model dependence comes from the $1^{-}$ contribution, which is the largest one due to the longer range of the dipole NP potential and which is almost entirely dominated by the giant dipole resonances.

\begin{table}[!tbp]
\begin{center}
\caption{Model dependence of contributions with different multipolarities to the Coulomb SE-NP shift (in meV) of the $1s_{1/2}$ level of hydrogenlike $^{208}\text{Pb}^{81+}$. The corrections are obtained by means of the formulas from Ref.~\cite{1996_Nefiodov} using the same experimental and EWSR nuclear parameters (last row) as well as utilizing the $B(EL)_{\lambda}$ and $\omega_{\lambda}$ values obtained from the~\texttt{skyrme\textunderscore rpa} program~\cite{skyrme_rpa} with eleven different Skyrme parametrizations.}
{\renewcommand{\arraystretch}{1.2}
\renewcommand{\tabcolsep}{0.3cm}
\begin{tabular}{lccccc}
\hline \hline
& $0^{+}$ & $1^{-}$ & $2^{+}$ & $3^{-}$ & Total \\
\hline
KDE0  & -3.2 & -18.8 & -4.8 & -1.8 & -28.7 \\
SKX   & -3.0 & -19.3 & -5.4 & -1.9 & -29.6 \\
SLy5  & -3.4 & -19.5 & -5.0 & -1.9 & -29.7 \\
BSk14 & -3.2 & -20.0 & -5.1 & -1.8 & -30.1 \\
SAMi  & -3.4 & -21.6 & -5.1 & -2.0 & -32.1 \\
NRAPR & -3.5 & -21.3 & -5.2 & -2.3 & -32.4 \\
SkP   & -3.6 & -19.9 & -5.6 & -1.9 & -31.0 \\
SkM*  & -3.6 & -21.6 & -5.3 & -1.9 & -32.5 \\
SGII  & -3.6 & -22.1 & -5.3 & -2.0 & -33.1 \\
SKI3  & -3.2 & -20.9 & -4.7 & -1.9 & -30.8 \\
LNS   & -3.2 & -19.5 & -4.7 & -1.6 & -29.0 \\
\hline
Exp.$+$EWSR & -3.3 & -17.2 & -5.8 & -2.6 & -28.9 \\
\hline \hline
\end{tabular}} 
\label{tab:bel}
\end{center}
\end{table}

The use of Eq.~\eqref{eq:NP_Nef} involves certain assumptions about the radial dependence of the Coulomb part of the NP tensor.
It is not uncommon to employ the same radial functions~\eqref{eq:F_L}, which correspond to harmonic surface vibrations, for all types of nuclear excitations, except for the special case of the breathing modes, where Eq.~\eqref{eq:F_0} is used instead.
This simplification is justified by the small overlap between electronic wavefunctions and a nucleus such that the details of this radial dependence are of minor importance.
Therefore, it is also of interest to test this approximation by using~$\Pi^{00}_{\text{N}}$ constructed directly from the RPA charge transition densities~$\rho^{\text{tr}}_{\lambda}$ (given by Eq.~\eqref{eq:td}) with a unique radial dependence for each nuclear excitation.
We compare the $\{B(EL)_{\lambda}+F_L\}$ and the $\rho^{\text{tr}}_{\lambda}$ results for the SLy5 parametrization in the first two rows of Table~\ref{tab:cnp}, again using the $1s_{1/2}$ state of~$^{208}\text{Pb}^{81+}$ as an example.
One can see that the more accurate and detailed treatment of the Coulomb SE-NP correction by means of $\rho^{\text{tr}}_{\lambda}$ leads to slightly larger values, with most of the difference coming from the dominant~$1^{-}$ contribution.
Nevertheless, the overall agreement between the EWSR estimation from the last row of Table~\ref{tab:bel} and the microscopic calculation from the second row of Table~\ref{tab:cnp} is noteworthy, given how much simpler the former is compared to the latter.

\subsection{Full effective self-energy} \label{se_full}
\begin{table}[!tbp]
\begin{center}
\caption{SE-NP contributions~(in meV) with different multipolarities for the $1s_{1/2}$ state of $^{208}\text{Pb}^{81+}$ obtained with the SLy5 parametrization of the Skyrme force. 
Two ways of implementing the Coulomb approximation are compared to the full calculation including the transverse electron-nucleus interaction. See Subsections~\ref{se_coulomb} and~\ref{se_full} for more details.}
{\renewcommand{\arraystretch}{1.2}
\renewcommand{\tabcolsep}{0.161cm}
\begin{tabular}{lccccc}
\hline \hline
[SLy5] & $0^{+}$ & $1^{-}$ & $2^{+}$ & $3^{-}$ & Total \\
\hline
Coulomb, $\{B(EL)_{\lambda}+F_L\}$    & -3.4 & -19.5 & -5.0 & -1.9 & -29.7 \\
Coulomb, $\rho^{\text{tr}}_{\lambda}$ & -3.6 & -22.3 & -5.9 & -2.2 & -34.0 \\
Coulomb$+$transverse                  & -3.5 & -30.1 & -5.9 & -2.2 & -41.8 \\
\hline \hline
\end{tabular}}
\label{tab:cnp}
\end{center}
\end{table}

\begin{figure*}[!htbp]%
\centering
\subfloat[\centering]{{\includegraphics[width=8.3cm]{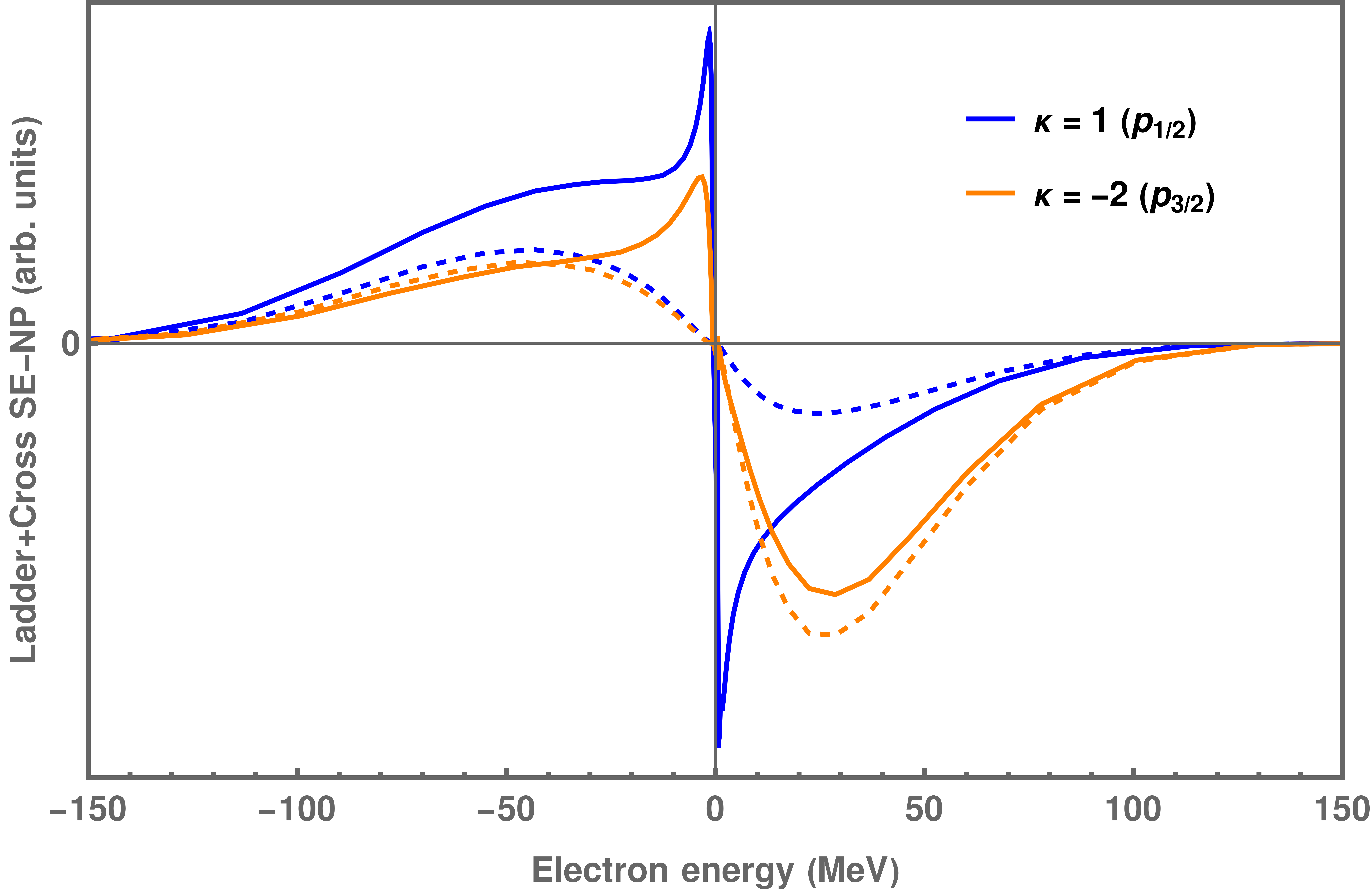} }}%
\qquad
\subfloat[\centering]{{\includegraphics[width=8.3cm]{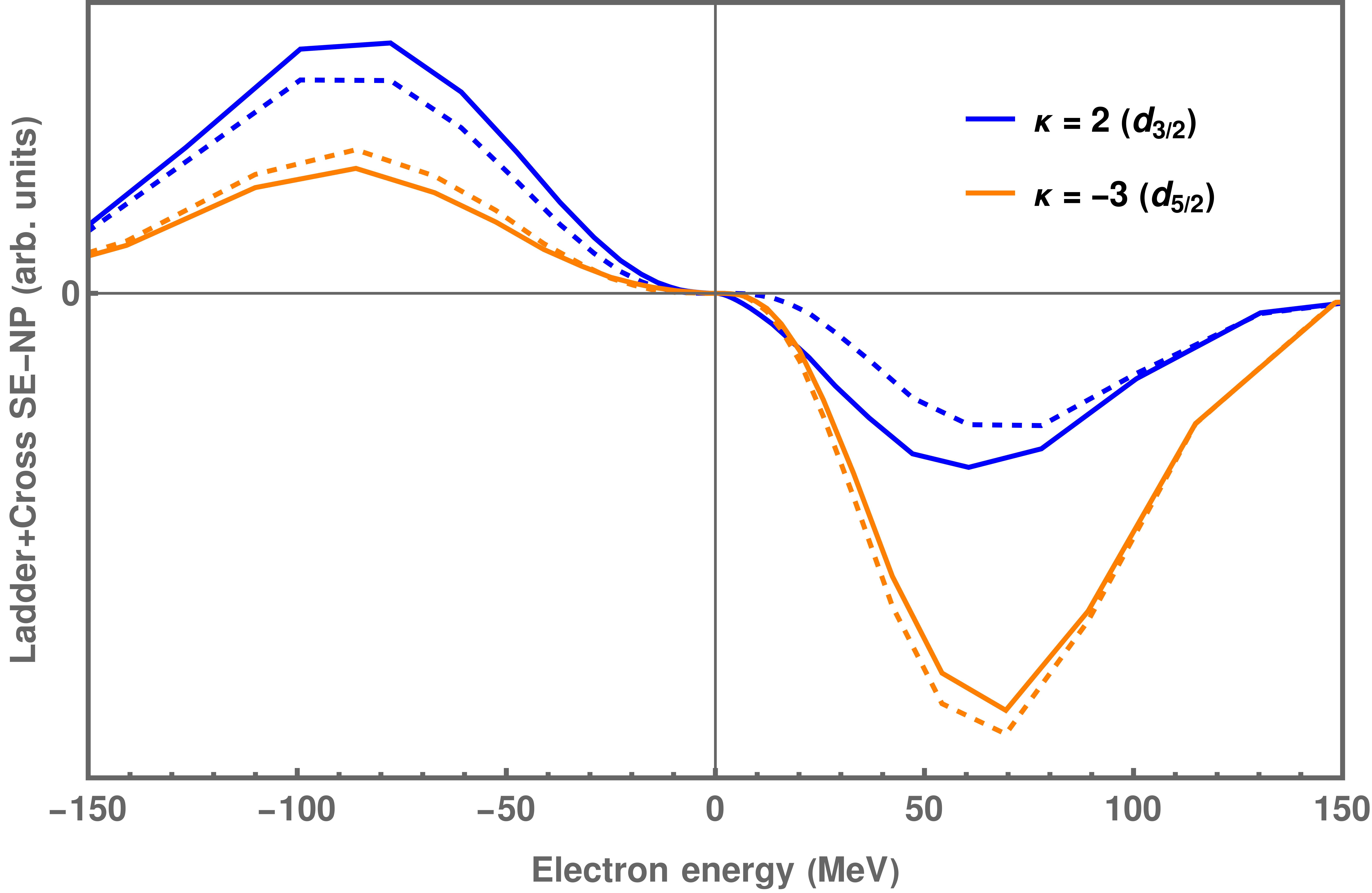} }}%
\vspace{-0.15cm}
\caption{Contributions from different intermediate electronic states to the sum of the ladder and the cross parts of the SE-NP correction to the $1s_{1/2}$ state of $^{208}\text{Pb}^{81+}$ for single fixed $1^{-}$~(a) and $2^{+}$~(b) nuclear excitations with the energies of 12.8~MeV and 12.2~MeV, respectively. 
Dashed lines represent the Coulomb approximation, while the solid lines correspond to the full electron-nucleus interaction including the transverse part.
Different colors are used for the two allowed values of the relativistic angular number~$\kappa$ of the electronic states in both cases.
The nuclear part of the calculations is performed using the SLy5 Skyrme parametrization.}%
\label{fig:npdensity}%
\vspace{-0.15cm}
\end{figure*}
The assumption of including only the~$\Pi^{00}_{\text{N}}$ component of the NP tensor turns out to be not as accurate as was initially thought.
It was argued in Refs.~\cite{Tanaka_1994, Haga_e_2002} that the transverse electron-nucleus interaction is not negligible due to the interference term with the Coulomb contribution.
This fact is illustrated in the last row of Table~\ref{tab:cnp}, where we present the SE-NP corrections with the full electromagnetic electron-nucleus interaction taken into account by means of the formulas from Ref.~\cite{Haga_e_2002}.
While the $0^{+}$, $2^{+}$, and $3^{-}$ contributions are not influenced much by the inclusion of the transverse part, the dominant $1^{-}$ correction is increased by as much as 35\%.

In order to gain more insight into this difference, we plot in Fig.~\ref{fig:npdensity} the sum of the ladder and the cross contributions as a function of the intermediate electron energy for fixed $1^{-}$~(a) and $2^{+}$~(b) nuclear excitations with the energies of 12.8~MeV and 12.2~MeV, respectively.
Both of these nuclear states represent the largest contributions from the giant resonances.
The dashed lines are used for the Coulomb approximation, while the solid lines correspond to the full electron-nucleus interaction.
Different colors represent the two allowed values of the relativistic angular quantum number~$\kappa$ of the intermediate electronic states in both cases.
First, it is worth pointing out the large cancellations that take place between the contributions from the positive- and negative-energy parts of the Dirac spectrum, which makes the precise evaluation of the total SE-NP correction a challenging task.
Then, by comparing Fig.~\ref{fig:npdensity}(a) and Fig.~\ref{fig:npdensity}(b) one notices a striking difference in how the inclusion of the transverse electron-nucleus interaction affects these two cases.
While it does not create much of a difference for the $2^{+}$ mode, sharp peaks appear in the low-energy region for the $1^{-}$ one.
This unique behaviour is due to the fact that only in the dipole case there are transverse form factors containing the non-vanishing spherical Bessel function $j_0$ at zero momentum transfer~\cite{Haga_e_2002}.
Here we note that the transverse interaction also leads to a small magnetic~$1^{+}$ correction of approximately +1.5~meV for the ground state of $^{208}\text{Pb}^{81+}$, thus contributing with the opposite sign compared to the electric excitation modes.

\begin{table*}[!tbp]
\begin{center}
\caption{Full SE-NP corrections (in meV) to the $1s_{1/2}$, $2s_{1/2}$, and $2p_{1/2}$ energy levels of hydrogenlike ions $^{208}\text{Pb}^{81+}$, $^{120}\text{Sn}^{49+}$, $^{90}\text{Zr}^{39+}$, $^{60}\text{Ni}^{27+}$, and $^{40}\text{Ca}^{19+}$.
The results are given for eleven different Skyrme parametrizations in order to explore nuclear model dependence.}
{\renewcommand{\arraystretch}{1.2}
\renewcommand{\tabcolsep}{0.177cm}
\begin{tabular*}{\textwidth}{l|ccc|ccc|ccc|ccc|ccc}
\hline \hline
\multirow{2}{*}{Model} & \multicolumn{3}{c|}{$^{208}\text{Pb}^{81+}$} & \multicolumn{3}{c|}{$^{120}\text{Sn}^{49+}$} & \multicolumn{3}{c|}{$^{90}\text{Zr}^{39+}$} & \multicolumn{3}{c|}{$^{60}\text{Ni}^{27+}[\times 10^1]$} & \multicolumn{3}{c}{$^{40}\text{Ca}^{19+}[\times 10^2]$} \\
& $1s_{1/2}$ & $2s_{1/2}$ & $2p_{1/2}$ & $1s_{1/2}$ & $2s_{1/2}$ & $2p_{1/2}$ & $1s_{1/2}$ & $2s_{1/2}$ & $2p_{1/2}$ & $1s_{1/2}$ & $2s_{1/2}$ & $2p_{1/2}$ & $1s_{1/2}$ & $2s_{1/2}$ & $2p_{1/2}$ \\
\hline
KDE0  & -42.5 & -6.6 & 1.1 & -2.5 & -0.29 & 0.076 & -0.87 & -0.098 & 0.020 & -2.1 & -0.24 & 0.022 & -5.6 & -0.64 & 0.017\\
SKX   & -44.3 & -6.8 & 1.2 & -2.6 & -0.31 & 0.084 & -0.97 & -0.108 & 0.023 & -2.4 & -0.27 & 0.025 & -6.8 & -0.79 & 0.022\\
SLy5  & -43.0 & -6.8 & 0.8 & -2.3 & -0.27 & 0.060 & -0.78 & -0.090 & 0.016 & -1.8 & -0.20 & 0.017 & -4.6 & -0.54 & 0.014\\
BSk14 & -43.8 & -6.9 & 0.9 & -2.4 & -0.28 & 0.071 & -0.85 & -0.096 & 0.019 & -2.0 & -0.23 & 0.020 & -5.2 & -0.60 & 0.016\\
SAMi  & -49.2 & -7.3 & 1.8 & -3.2 & -0.36 & 0.115 & -1.21 & -0.133 & 0.031 & -3.0 & -0.34 & 0.033 & -8.6 & -0.98 & 0.028\\
NRAPR & -50.9 & -7.4 & 2.2 & -3.5 & -0.39 & 0.140 & -1.39 & -0.151 & 0.037 & -3.6 & -0.40 & 0.040 &-10.5 & -1.20 & 0.035\\
SkP   & -45.8 & -7.1 & 1.2 & -2.6 & -0.31 & 0.082 & -0.95 & -0.107 & 0.022 & -2.3 & -0.26 & 0.024 & -6.1 & -0.70 & 0.019\\
SkM*  & -49.5 & -7.3 & 1.9 & -3.3 & -0.37 & 0.119 & -1.23 & -0.136 & 0.032 & -3.2 & -0.36 & 0.035 & -8.7 & -0.99 & 0.028\\
SGII  & -50.1 & -7.5 & 1.8 & -3.2 & -0.37 & 0.114 & -1.22 & -0.134 & 0.031 & -3.1 & -0.35 & 0.034 & -8.6 & -0.99 & 0.028\\
SKI3  & -44.0 & -7.0 & 0.7 & -2.3 & -0.28 & 0.060 & -0.79 & -0.090 & 0.017 & -1.7 & -0.20 & 0.017 & -4.7 & -0.54 & 0.014\\
LNS   & -45.5 & -6.9 & 1.3 & -2.8 & -0.32 & 0.094 & -1.02 & -0.113 & 0.025 & $-$  & $-$   & $-$   & -6.9 & -0.80 & 0.022\\
\hline \hline
\end{tabular*}} 
\label{tab:senp}
\end{center}
\end{table*}

We now present in Table~\ref{tab:senp} the results of our full SE\nobreakdash-NP calculations including the $0^{+}$, $1^{-}$, $2^{+}$, $3^{-}$, $4^{+}$, $5^{-}$, and $1^{+}$ excitation modes.
The energy shifts are given for five hydrogenlike ions, namely $^{208}\text{Pb}^{81+}$, $^{120}\text{Sn}^{49+}$, $^{90}\text{Zr}^{39+}$, $^{60}\text{Ni}^{27+}$, and $^{40}\text{Ca}^{19+}$, in the $1s_{1/2}$, $2s_{1/2}$, and $2p_{1/2}$ states.
Again, in order to explore the nuclear model dependence, the values have been obtained using eleven different Skyrme parametrizations, except for the case of $^{60}\text{Ni}^{27+}$ where the LNS calculation breaks down.
In general, it can be seen that the SE-NP correction exhibits a rather steep decrease for lower charge numbers~$Z$ and loses about an order of magnitude when going to the next principal or angular quantum number.

A conspicuous feature of Table~\ref{tab:senp} is that the inclusion of the transverse electron-nucleus interaction makes the SE-NP corrections to the $2p_{1/2}$ state positive.
As with the $1s_{1/2}$ correction discussed above, the transverse interaction mainly affects the dominant $1^{-}$ contribution also in this case, which would otherwise be negative in the Coulomb approximation.
Moreover, the sum of the ladder and the cross diagrams for the $2p_{1/2}$ state is still negative in the full calculation, and it is the seagull term that plays the crucial role in making the total $1^{-}$ contribution positive.
This behaviour has been confirmed by performing the calculations in both Coulomb and Feynman gauges, which lead to quite different contributions from the individual diagrams but eventually produce the same total value.
To the best of our knowledge, this unusual result has not been reported in the literature so far, and its exact physical interpretation is rather elusive.

Finally, when it comes to high-precision tests of QED, one of the most important aspects of the calculations is to quantify the theoretical uncertainty.
In the case of the NP correction, this uncertainty predominantly stems from the fact that an \textit{ab initio} computation of the entire nuclear spectrum is not feasible such that an effective model has to be applied instead.
The results in Table~\ref{tab:senp} for a wide range of Skyrme parametrizations provide a good sense of the nuclear model dependence from the microscopic point of view.
The general trend is that smaller corrections are associated with a larger spread of the theoretical predictions, with the uncertainties (taken as half the range divided by the mean value) reaching up to 40\nobreakdash--50\% for $^{40}\text{Ca}^{19+}$.
Similar high uncertainties are also observed in the case of the $2p_{1/2}$ state for all of the ions considered.
The smallest spread, and thus the best precision, is found for the $1s_{1/2}$ and $2s_{1/2}$ levels of the heaviest $^{208}\text{Pb}^{81+}$ ion, where the theoretical uncertainties are below 10\%.
Such a degree of control over the nuclear-structure effects makes this system an even more promising platform for testing QED in the strongest electromagnetic fields.
It is also interesting to note that in most cases in Table~\ref{tab:senp} the NRAPR parametrization represents significant outliers, while such a behaviour is not observed for muonic atoms~\cite{2022_Valuev}.
Even though there is no strong physical argument either in favor or against the reliability of a particular model, it seems nevertheless likely that the NRAPR parametrization tends to overestimate the magnitudes of the NP effect in electronic systems, thereby providing conservative upper bounds.

\section{Conclusions and outlook}
In this work we have presented the field-theoretical formalism for the NP effect as well as our full leading-order computational results involving a detailed microscopic nuclear description.
In general, the formalism predicts two distinct NP contributions of order $\alpha^2$.
One of them, which is represented by the effective vacuum-polarization diagram, has been somewhat of a mystery in the literature.
On the one hand, it was suggested in Ref.~\cite{1996_Labzowsky} that a part of this correction could be distinguished from the ordinary FNS effect due to its different scaling with the nuclear charge radius.
On the other hand, it still remained unclear how to extract such a contribution, and this VP\nobreakdash-NP correction has been largely ignored in NP calculations, with the only exceptions of Refs.~\cite{Haga_mu_2002, Haga_mu_2007}.
It is especially important to resolve this ambiguity, as inclusion of the VP\nobreakdash-NP term would modify the total NP corrections significantly, e.g., by canceling more than half of the usual SE\nobreakdash-NP contribution for the $1s_{1/2}$ state of $^{208}\text{Pb}^{81+}$.
By considering the effect of the induced vacuum polarization on the nuclear size, we have demonstrated that the VP\nobreakdash-NP contribution is already contained in the standard FNS correction. 
Importantly, we have checked that this result holds for both electronic and muonic systems, as the latter are used to experimentally measure nuclear charge radii.
Consequently, we conclude that the VP\nobreakdash-NP term should indeed be omitted in order to avoid double counting. 

The other NP contribution of order $\alpha^2$ is given by the effective self-energy diagram, which can also be interpreted as the two-photon exchange between the bound electron and the nucleus.
This correction is often evaluated by neglecting the effects of the nuclear three-currents, resulting in what we call the Coulomb approximation.
In addition, the crucial contributions from the giant nuclear resonances are usually estimated by means of phenomenological EWSR.
While we have found this approach to be in remarkably good agreement with our more detailed microscopic calculations, we have also demonstrated the importance of including the transverse part of the electron-nucleus interaction, which plays an essential role for an accurate evaluation of the dipole part of the SE\nobreakdash-NP correction.
For instance, the transverse interaction in $^{208}\text{Pb}^{81+}$ increases the dipole contribution to the $1s_{1/2}$ energy shift by as much as 35\% and may even lead to a sign change from negative to positive in the case of the $2p_{1/2}$ state as suggested by our results.
Therefore, we emphasize that the commonly used Coulomb approximation as well as the corresponding effective potentials~\cite{2002_Nefiodov, 2014_Volotka, 2021_Flambaum} significantly underestimate the magnitude of the SE\nobreakdash-NP correction, and full calculations are needed in~order to deliver satisfactory accuracy.

Finally, we have extensively analyzed the nuclear model dependence of our NP calculations, allowing us to reliably estimate the dominant theoretical uncertainty.
The smallest spread of the results for eleven different models has been observed for the heaviest $^{208}\text{Pb}^{81+}$ ion, with the uncertainties being below 10\% for the $1s_{1/2}$ and $2s_{1/2}$ states.
From one point of view, this level of precision for nuclear-structure effects makes $^{208}\text{Pb}^{81+}$ an appealing platform for stringent tests of QED in extreme electromagnetic fields.
Alternatively, a set of different predictions from various nuclear models in combination with high enough experimental precision may even potentially offer an opportunity to test and improve our understanding of the inner workings of the nucleus itself.

\begin{acknowledgments}
The authors wish to thank Gianluca Col\`{o} and Xavier Roca-Maza for their help with the \verb|skyrme_rpa| code as well as Vladimir A. Yerokhin for insightful discussions.
\end{acknowledgments}

%\bibliography{refs}

\end{document}